\documentclass[cameraready]{Interspeech}
\usepackage{multirow}
\usepackage[table]{xcolor}
\newcommand{\gray}{\cellcolor{black!15}}
\usepackage{threeparttable}
 \usepackage[sort,compress]{cite}

\title{Low-Burden Data Augmentation for Dysarthric ASR via Zero-Shot \\Voice Cloning}

\author[affiliation={1}, orcid=0000-0001-9032-6781]{Satwinder}{Singh}
\author[affiliation={1}, orcid=0009-0006-8744-5898]{Qianli}{Wang}
\author[affiliation={1}, orcid=0009-0009-9809-4589]{Zihan}{Zhong}
\author[affiliation={2}, orcid=0000-0003-4207-748X]{Clarion}{Mendes}
\author[affiliation={2}, orcid=0000-0002-5631-2893]{Mark}{Hasegawa-Johnson}
\author[affiliation={1}, orcid=0000-0002-1812-4285]{Waleed}{Abdulla}
\author[affiliation={1}, orcid=0000-0003-1543-5931]{Seyed Reza}{Shahamiri}
\address{
   $^1$ DeepNet Discovery Network, University of Auckland, New Zealand \\
   $^2$ University of Illinois Urbana-Champaign, USA
}
\email{admin@profsatwinder.com, \{qwan121,zzho680\}@aucklanduni.ac.nz, \{cmendes2, jhasegaw\}@illinois.edu, w.abdulla@auckland.ac.nz, admin@rezanet.com}


\begin{document}

\maketitle

\begin{abstract}
Automatic speech recognition remains unreliable for dysarthric speech due to data scarcity and high inter-speaker variability. While synthetic data can address these gaps, traditional methods often require extensive speaker-specific data, reintroducing the collection bottleneck. We investigate zero-shot voice cloning as a low-burden augmentation strategy, using Higgs Audio V2 to clone speakers in the TORGO dataset. We fine-tune (FT) Whisper-medium on cloned, real, and hybrid data and evaluate on held-out real speech. Compared to the zero-shot (31.62\%), Clone FT achieved a competitive 26.00\% WER, nearly matching the 24.44\% and 25.12\% seen with Real and Hybrid FT, respectively. Notably, Clone and Hybrid FT outperform Real FT for moderate-severe speakers. Clone FT achieves the best results (11.45\% relative) in cross-corpus evaluation on the SAP-1102. These results suggest that zero-shot cloning provides scalable training data that circumvents the costly data collection bottleneck. 
\end{abstract}

\section{Introduction}
Automatic speech recognition (ASR) has achieved strong accuracy on typical speech, driven by large-scale datasets and transformer-based models such as Whisper \cite{radford2023robust} and wav2vec 2.0 \cite{baevski2020wav2vec}. However, these gains do not extend reliably to dysarthric speech, which arises from neurological conditions such as Parkinson’s disease (PD), cerebral palsy (CP), and amyotrophic lateral sclerosis (ALS) \cite{rudzicz2012torgo}. Dysarthria introduces systematic distortions in articulation, timing, and phonation, leading to high inter-speaker variability across etiologies and substantial intra-speaker variability due to fatigue or disease progression \cite{kim2010frequency, enderby2013disorders}. As a result, even state-of-the-art ASR systems exhibit large performance degradation for moderate-to-severe speakers \cite{singh2024comprehensive, singh2025robust, zhong2025convolution}.

A central barrier to improving dysarthric ASR is data scarcity \cite{zheng2023improving, wang2025dysarthric}. Collecting dysarthric corpora is slow and expensive: recruitment is difficult, speakers fatigue quickly, transcription is labor-intensive, and clinical labels (e.g., severity, intelligibility) often require expert time \cite{yue2025challenges, hirsch2022reliability}. Benchmark datasets such as UASpeech \cite{kim2008dysarthric} and TORGO \cite{rudzicz2012torgo} remain small (\textless 20 dysarthric speakers), and even large initiatives such as the Speech Accessibility Project (SAP) \cite{hasegawa2024community} cannot match the scale of typical-speech training data. This structural scarcity limits both model robustness and personalization.

Synthetic speech is an appealing way to expand training data without repeatedly burdening dysarthric speakers. Prior work has explored text-to-speech (TTS) or voice conversion for adaptation, for example, by conditioning synthesis on speaker embeddings such as x-vectors \cite{wagner2025personalized}. More recent zero-shot voice cloning systems (e.g., VALL-E \cite{chen2024vall}, F5-TTS \cite{chen2024f5}, Higgs Audio \cite{higgsaudio2025}) can generate speech in a target voice from only seconds of reference audio, without speaker-specific fine-tuning. Yet for dysarthria, it remains unclear whether cloned speech provides a useful training signal for ASR: synthesis models trained largely on typical speech may smooth dysarthria-relevant cues (irregular timing, reduced coarticulation, unstable phonation) or introduce artifacts that bias fine-tuning \cite{soleymanpour2024accurate, huang2022towards}.

In this work, we conduct a controlled study to test whether zero-shot voice cloning can serve as \emph{low-burden} augmentation for dysarthric ASR. Using Higgs Audio V2 \cite{higgsaudio2025}, we clone dysarthric speakers present in the TORGO dataset from a \emph{single} reference utterance per speaker (average 7.2 seconds) and synthesize a training dataset (i.e., TORGO-Synth) from linguistically diverse, out-of-domain text prompts. We fine-tune Whisper-medium on cloned speech and evaluate exclusively on held-out \emph{real} TORGO test utterances under a strict lexical separation between cloning prompts and test transcripts. Beyond overall WER, we analyze (i) speaker similarity between clone and reference audio in a verifier embedding space and (ii) how ASR performance scales with the amount of synthetic data, revealing when cloning helps and when it can fail, and (iii) cross-corpus transfer on SAP-1102 \cite{hasegawa2024community}, assessing whether clone-based adaptation generalizes beyond the source dataset. 

Our results demonstrate that cloned speech provides a robust adaptation signal: Clone FT reduces TORGO WER from 31.62\% to 26.00\% (17.8\% relative) and identifies a non-monotonic scaling ``sweet spot" at 15 hours of synthetic data. Furthermore, the approach demonstrates strong generalization, improving cross-corpus performance on SAP-1102 from 14.50\% to 12.84\% (11.45\% relative).

\vspace{-5pt}
\section{Related Work}
\subsection{Synthetic Data for Dysarthric ASR}
Data scarcity has motivated a range of augmentation strategies for dysarthric ASR. Conventional signal-level perturbations (e.g., speed or noise) can increase acoustic diversity but do not create new lexical content or explicitly model dysarthria-specific timing and articulation patterns \cite{park2019specaugment, ko2015audio, lei2023phaseperturbation}. Consequently, recent work has explored speech generation to expand training corpora without requiring repeated recording sessions. TTS has been investigated for producing dysarthric-like speech \cite{leung2024training, soleymanpour2024accurate, hermann2023few}, and personalized setups have also been explored that condition synthesis on speaker representations (e.g., x-vectors) to support adaptation \cite{wagner2025personalized}. These approaches can reduce the transcription burden and increase lexical coverage, but many still require speaker-specific fine-tuning or multiple enrollment utterances, which partially reintroduces the data-collection bottleneck. In addition, because dysarthric speech differs systematically from typical speech in prosody, pausing, and phonation, TTS models trained predominantly on typical speech may risk normalizing these atypical cues unless dysarthria-specific controls are introduced, limiting how faithfully they reflect dysarthric acoustics \cite{soleymanpour2024accurate, huang2022towards}. 

Voice conversion (VC) has also been used to transform typical speech into dysarthric-like speech \cite{zheng2023improving, huang2022towards, vachhani2018data, jiao2018simulating}. VC methods often rely on parallel or carefully matched recordings and may require pathology-specific mappings that do not scale across speakers and etiologies. Moreover, both TTS and VC pipelines can be constrained by the linguistic coverage of available prompts or paired material, reducing their value as general-purpose training data for open-vocabulary ASR.

\vspace{-5pt}
\subsection{Zero-Shot Voice Cloning}
Zero-shot voice cloning offers a particularly low-burden route to synthetic data generation: modern models can generate speech in a target voice from only seconds of reference audio and arbitrary text prompts, without speaker-specific fine-tuning \cite{chen2024vall, casanova2022yourtts, chen2024f5, higgsaudio2025}. This capability enables out-of-vocabulary synthesis and can substantially expand lexical coverage without collecting additional dysarthric recordings. Prior work has argued that exposing ASR to synthetic renditions of novel phrases can improve recognition on unseen lexical items \cite{harvill2021synthesis}. However, for dysarthria, it remains an open question whether zero-shot cloning preserves the speaker-discriminative and dysarthria-relevant cues needed to provide a reliable adaptation signal, or whether synthesis artifacts and prosody normalization can instead bias fine-tuning.

We address these open questions by providing a controlled evaluation of a single-utterance enrollment voice cloning for dysarthric ASR. Unlike prior work that requires multi-utterance enrollment, we adopt Higgs Audio V2 to synthesize speech from a single reference per TORGO speaker. Our study uniquely combines (i) a speaker similarity analysis, (ii) a data scaling analysis, and (iii) a cross-corpus transfer evaluation on SAP-1102, providing a comprehensive assessment of how zero-shot cloning impacts downstream ASR robustness.

\vspace{-3pt}
\section{Method}
\subsection{Voice Cloning Setup}
In this study, we adopted the Higgs Audio V2 voice cloning model \cite{higgsaudio2025} to generate synthetic dysarthric speech samples. It is a large-scale (5B parameters), open-source state-of-the-art audio foundation model trained on over 10 million hours of diverse audio-text pairs. The model performs zero-shot TTS synthesis without requiring speaker-specific fine-tuning or post-training adaptation. The architecture employs a dual feed-forward network audio adapter that processes acoustic (prosodic/timbral) representations, enabling the modeling of vocal characteristics. The model supports a system prompt to control the lexical and stylistic variation through the semantic encoder. It maintains temporal coherence across long-form synthesis, ensuring stable prosody and speaker consistency. These properties make it well-suited for low-resource pathological speech applications, where preserving atypical timing, phonation, and speaker-specific traits is critical for downstream tasks like ASR.

\begin{figure}
    \centering
    \includegraphics[width=1\linewidth]{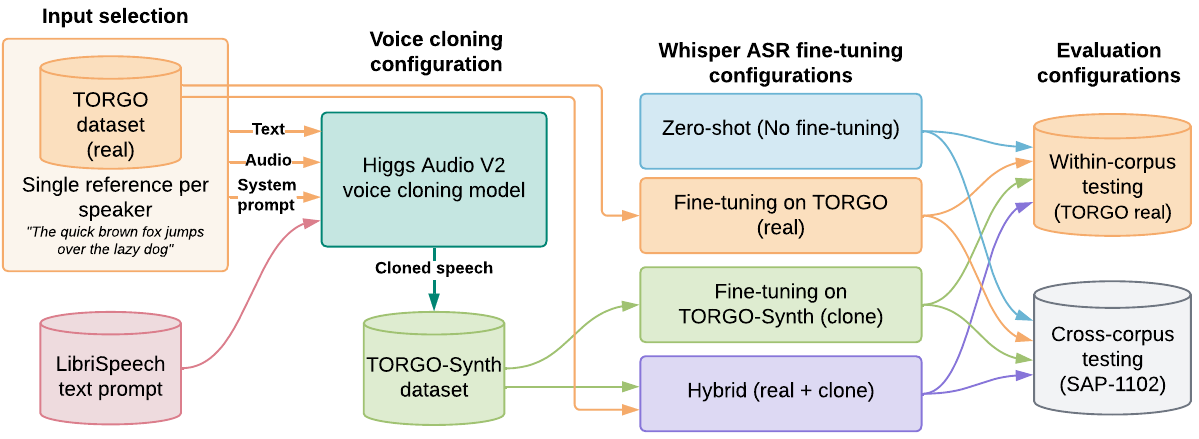}
    \caption{Overview of voice cloning and ASR fine-tuning and evaluation pipeline.}
    \label{fig:framework}
    \vspace{-15pt}
\end{figure}

Figure \ref{fig:framework} shows our overall pipeline. For each TORGO speaker \cite{rudzicz2012torgo}, we selected a single reference utterance averaging 7.2 seconds. This duration aligns with the 5--10 second range recommended for optimal zero-shot synthesis with Higgs Audio V2 \cite{higgsaudio2025}. We chose the phonetically rich sentence ``\textit{The quick brown fox jumps over the lazy dog}" as the reference. Additionally, we provide out-of-domain text prompts from the LibriSpeech 100h dataset \cite{panayotov2015librispeech} for the model to clone. We used a minimal system prompt, “\textit{Generate audio following instruction}”, without any scene description or additional prompt-based conditioning, so that generation relied primarily on the enrollment transcript and reference audio. This was important for dysarthric speech cloning because additional prompt-level conditioning could impose unintended prosodic or acoustic patterns, potentially reducing the fidelity with which speaker-specific dysarthric traits were preserved.
To balance naturalness and expressive variability during cloning, we used a sampling configuration with temperature 1.0, \texttt{top\_k} 50, and \texttt{top\_p} 0.95.

\vspace{-5pt}
\subsection{The Whisper Model for ASR Task}
For the evaluation of cloned speech, we employed the Whisper model \cite{radford2023robust}.  Following our previous work \cite{singh2024comprehensive,singh2025robust}, we employed the multilingual Whisper-medium model (769M parameters) as our ASR backbone. The Whisper model adopts an encoder-decoder Transformer architecture tailored for large-scale speech recognition. The encoder processes 80 log-Mel spectrogram features extracted from raw audio, using a series of multi-head self-attention layers to capture both local acoustic patterns and long-range temporal dependencies. The decoder, conditioned on the encoder's hidden states, autoregressively generates text tokens via cross-attention layers that link the acoustic and linguistic representations.  Whisper is trained on a multitask objective that includes speech recognition, translation, and language identification. This enables the decoder to handle multilingual inputs and adapt to diverse tasks within the same architecture. Unlike conventional ASR pipelines that require separate acoustic and language models, Whisper integrates these components into a single end-to-end framework. Fine-tuning was performed with an effective batch size of 32, a learning rate of 5e-6, and weight decay of 0.01. For decoding, we used beam search with a beam size of 10 and a \texttt{no\_repeat\_ngram\_size} of 3 to discourage repetitive outputs.

\vspace{-6pt}
\section{Experimental Setup}
\subsection{Datasets}
\label{sec:datasets}
\textbf{TORGO dataset} \cite{rudzicz2012torgo}, developed by the University of Toronto, provides clinically annotated speech recordings from individuals with dysarthria due to CP and ALS. The dataset contains approximately 23 hours of speech from 8 dysarthric speakers (5 male, 3 female; 7 with CP, 1 with ALS) and 7 neurotypical controls. Recordings include non-words (e.g., /iy-p-ah/), isolated words (e.g., ``yes", ``no"), phonetically balanced restricted sentences, and unrestricted spontaneous speech, offering rich phonetic and prosodic coverage. Dysarthric speakers are clinically categorized by severity: severe, moderate-severe, moderate, and mild. For this study, we focus exclusively on restricted speech utterances to ensure linguistic complexity and natural prosody for robust evaluation of zero-shot voice cloning under realistic conditions.

\textbf{TORGO-Synth dataset.} We refer to our synthesized dataset as TORGO-Synth.
We used the LibriSpeech 100h dataset \cite{panayotov2015librispeech} as a source of text prompts for voice cloning. Prompts were extracted as plain transcripts, duplicates were removed, and the remaining sentences were filtered to retain those with 3 to 20 words. The resulting text prompts were then randomly shuffled to remove sequential or narrative structure from the original audiobooks and reduce ordering bias during synthesis. To ensure strict lexical separation, any LibriSpeech prompts overlapping with the TORGO and SAP-1102 datasets were removed, preventing lexical leakage during training.

\begin{figure}[t]
    \centering
    \includegraphics[width=1\linewidth]{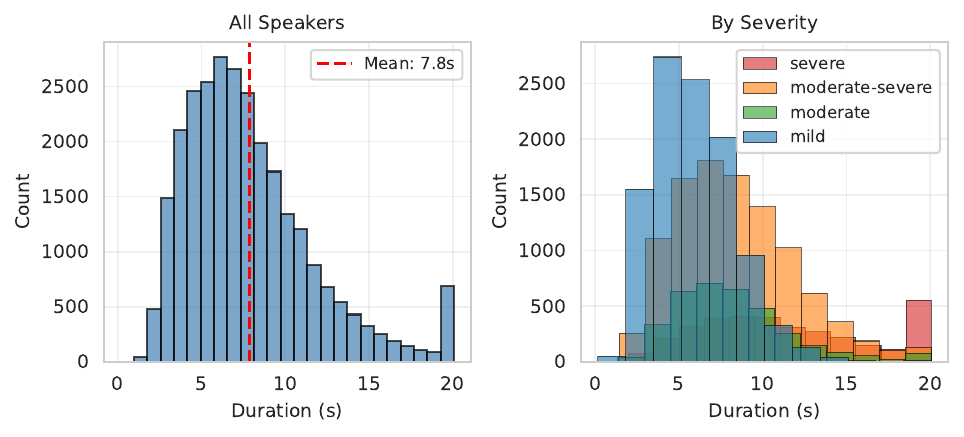}
    \caption{Duration distribution of cloned utterances in TORGO-Synth.}
    \label{fig:stat}
    \vspace{-15pt}
\end{figure}

TORGO-Synth consists of 8,289 utterances totaling 18 hours (Train: 15 h, Validation: 3 h), encompassing eight speakers across the full spectrum of dysarthric severity (severe, moderate-severe, moderate, and mild). As shown in Figure~\ref{fig:stat}, the dataset exhibits a distinctive long-tail duration distribution with a mean of 7.8 s ($\sigma = 3.84$ s) and utterances extending up to 20 s. This profile is primarily driven by the severe and moderate-severe cohorts, M04 (3.259 h), F01 (2.229 h), M01 (2.353 h), and M02 (2.512 h), providing a clinically authentic representation of the reduced speaking rates characteristic of significant motor impairment. The dataset is complemented by moderate (M05: 2.487 h) and mild speakers (F03: 1.433 h, F04: 1.958 h, M03: 1.772 h) to ensure broad acoustic coverage. Representative audio samples are available online\footnote{Audio samples: \url{https://ai-research-submissions.github.io/interspeech_audio_samples/}}.

\textbf{SAP-1102 dataset}. For the cross-corpus generalization analysis, we utilized the SAP-1102 dataset \cite{hasegawa2024community}, produced by the University of Illinois Urbana-Champaign (November 02, 2025 release). As the test set is reserved for the upcoming challenge, we carefully sampled a test subset of 500 utterances from the development set, encompassing ALS, Cerebral Palsy, and Parkinson’s disease etiologies, and covering 58 unique speakers. We only chose \textit{Novel Sentences}, which are read-speech samples from novels in the SAP-1102 dataset.

\begin{figure}[t]
    \centering
    \includegraphics[width=0.9\linewidth]{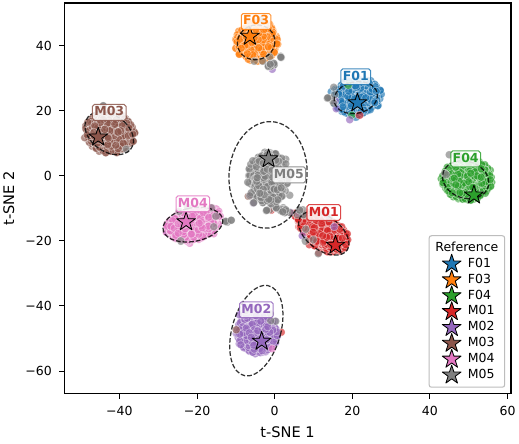}
    \caption{t-Distributed Stochastic Neighbor Embedding (t-SNE) projection of TORGO speaker embeddings. Stars ($\star$) and dots ($\bullet$) denote original and cloned speech, respectively. Dashed ellipses represent the 95\% confidence regions ($2\sigma$) for each speaker distribution.}
    \label{fig:tsne}
    \vspace{-15pt}
\end{figure}

\vspace{-5pt}
\subsection{Experimental Configurations}
We evaluate four experimental configurations. The \textit{Zero-Shot} configuration uses a pretrained Whisper-medium model without any fine-tuning, serving as our baseline. The \textit{Real} configuration fine-tunes the model exclusively on real speech samples from the TORGO dataset. The \textit{Clone} configuration fine-tunes the model solely on synthetic speech from the TORGO-Synth dataset. Finally, the \textit{Hybrid} configuration fine-tunes the model on a combination of both real and synthetic speech from TORGO and TORGO-Synth, respectively. In all configurations, models are evaluated on a real speech test set from the TORGO dataset, held out from all training and validation splits. Additionally, to assess cross-corpus generalization, all four configurations are evaluated on an additional test set sampled from the development split of the SAP-1102 dataset, as described in Section~\ref{sec:datasets}.

\vspace{-5pt}
\subsection{Evaluation Metrics}
ASR performance is reported using Word Error Rate (WER). To evaluate the acoustic similarity between cloned and real speech, we visualize speaker embeddings using t-SNE, extracted with NVIDIA's TitaNet speaker verification model \cite{koluguri2022titanet}, and measure similarity using cosine distance.

\section{Results and Discussion}

\subsection{Speaker Similarity Analysis}
Figure \ref{fig:tsne}  shows 2D t-SNE projections of TitaNet speaker embeddings \cite{koluguri2022titanet} for cloned utterances, with the single enrollment reference per speaker marked by a star. Several speakers (e.g., F03, M03, F01, and F04) form tight, well-separated clusters, with the reference embedded within the same region, indicating strong speaker consistency in the verifier embedding space. In contrast, M05 exhibits a substantially larger, more diffuse cluster near the center of the space, increasing proximity to other speakers and elevating the risk of confusion. M02 remains separated from other speakers but forms an elongated cluster, suggesting higher within-speaker variability; a subset of M02 clones drifts farther from the reference, consistent with moderately harder verification than for the compact clusters.

\vspace{-5pt}
\subsection{Downstream ASR Task Performance}
The experimental results (Table \ref{tab:wer_comparison}) demonstrate that all fine-tuning configurations significantly improve speech recognition for dysarthric speakers compared to the zero-shot baseline. Overall, fine-tuning on real dysarthric data achieved the lowest total WER of 24.44\%, representing an absolute Word Error Rate Reduction (WERR) of 7.18 percentage points (pp) and a relative WERR of 22.7\%. However, a granular severity-based analysis highlights the distinct advantage of using synthetic clone data, particularly for the most challenging speakers. In the Moderate-Severe group, both the Clone (39.95\% WER) and Hybrid (37.49\% WER) configurations outperformed the Real configuration (42.19\% WER), with the Hybrid approach yielding a substantial relative WERR of 31.4\% for this cohort. This highlights that synthetic clones can effectively augment the training distribution by providing high-fidelity, phonetically diverse samples that compensate for the scarcity and high variance of authentic dysarthric recordings. While the benefit of synthetic data diminishes for Mild speakers where the baseline is already near-ceiling, its ability to regularize performance for more impaired speakers suggests it is a critical tool for developing robust, personalized ASR systems. Notably, although Speaker M05 proved to be a zero-shot outlier, the overall improvements remain statistically significant ($p < $ 0.05), with the 95\% confidence interval for the Hybrid model [1.58, 12.94] confirming a reliable performance floor for synthetic-aided adaptation.

\begin{table}[t]
\centering
\caption{Speaker-level WER (\%) for Whisper-medium across zero-shot and fine-tuning configurations. \textbf{Bold} and \underline{underline} denote best and second-best results, respectively. $\Delta$WER and 95\% CIs indicate absolute reduction relative to zero-shot via paired bootstrap testing (speaker-level resampling, $n=8, B=10{,}000$).}
\vspace{-8pt}
\label{tab:wer_comparison}
\footnotesize
\setlength{\tabcolsep}{3.5pt}
\renewcommand{\arraystretch}{1.05}
\resizebox{\columnwidth}{!}{%
\begin{tabular}{lccccc}
\toprule
\textbf{Severity} & \textbf{Speaker} & \textbf{Zero-shot} & \textbf{Real} & \textbf{Clone} & \textbf{Hybrid} \\
\midrule\midrule
Severe & M04 & 82.32 & \textbf{60.22} & 63.54 & \underline{62.43} \\
\midrule
\multirow{3}{*}{Moderate-Severe} & F01 & 76.67 & 63.33 & \underline{48.33} & \textbf{43.33} \\
 & M01 & 45.28 & \textbf{27.83} & \underline{34.43} & 35.38 \\
 & M02 & 42.08 & \underline{35.42} & 37.08 & \textbf{33.75} \\
 & \gray \textit{Average} & \gray 54.68 & \gray 42.19 & \gray \underline{39.95} & \gray \textbf{37.49} \\
\midrule
Moderate & M05 & \textbf{23.81} & \underline{26.98} & 33.33 & 30.16 \\
\midrule
\multirow{3}{*}{Mild} & F03 & 15.56 & \textbf{14.60} & \textbf{14.60} & \underline{15.24} \\
 & F04 & \underline{2.65} & \textbf{1.59} & 3.17 & \underline{2.65} \\
 & M03 & 2.76 & \textbf{1.84} & \underline{2.30} & \textbf{1.84} \\
 & \gray \textit{Average} & \gray 6.99 & \gray \textbf{6.01} & \gray 6.69 & \gray \underline{6.58} \\
\midrule
Overall  WER        &  & 31.62 & \textbf{24.44} & 26.00 & \underline{25.12} \\
$\Delta$WER (pp) &  & --    & 7.18 & 5.62 & 6.50 \\
95\% CI          &  & --    & [1.95, 13.71] & [1.18, 11.60] & [1.58, 12.94] \\
\bottomrule
\end{tabular}
}
\vspace{-16pt}
\end{table}

\begin{table}[b]
\vspace{-10pt}
\centering
\caption{Average WER (\%) per severity group across fine-tuning data quantities (0h = zero-shot baseline; 5h--50h = hours of cloned speech used for fine-tuning.}
\label{tab:wer_severity}
\vspace{-8pt}
\renewcommand{\arraystretch}{1.1}
\setlength{\tabcolsep}{4pt}
\resizebox{\columnwidth}{!}{%
\begin{tabular}{lccccccccc}
\toprule
\textbf{Severity} & \textbf{0h} & \textbf{5h} & \textbf{10h} & \textbf{15h} & \textbf{20h} & \textbf{25h} & \textbf{30h} & \textbf{40h} & \textbf{50h} \\
\midrule\midrule
Severe          & 82.32 & 71.82 & 67.40 & \underline{63.54} & \textbf{62.43} & 64.64 & 69.61 & 67.40 & 64.09 \\
Moderate-Severe & 54.68 & 53.12 & 42.27 & \textbf{39.95} & 42.55 & 45.06 & \underline{40.55} & 41.30 & 46.25 \\
Moderate        & \textbf{23.81} & 34.92 & 38.10 & \underline{33.33} & 36.51 & 36.51 & 36.51 & 34.92 & 36.51 \\
Mild            &  6.99 &  7.62 &  6.74 &  \underline{6.69} &  \textbf{6.63} &  7.21 &  7.51 &  8.28 &  8.30 \\
\midrule
Overall WER        & 31.62 & 30.81 & 27.69 & \textbf{26.00} & \underline{26.47} & 28.23 & 27.22 & 28.37 & 28.37 \\
\bottomrule
\end{tabular}%
}
\vspace{-15pt}
\end{table}

\vspace{-5pt}
\subsection{Clone Data Scaling Results}
\vspace{-4pt}
To determine the optimal volume of synthetic augmentation, we conducted a scaling experiment as presented in Table \ref{tab:wer_severity}. The results reveal a non-monotonic relationship between synthetic data volume and performance, identifying an optimal ``sweet spot" at 15 hours of clone speech. The best overall performance is achieved at the 15-hour mark with a WER of 26.00\%, representing a relative WERR of 17.8\% over the zero-shot baseline. This advantage is most pronounced in the Severe and Moderate-Severe cohorts, where the Clone data provides a critical phonetic scaffold, reducing WER from 82.32\% to a minimum of 62.43\% (24.2\% WERR). However, increasing the data volume beyond 20 hours results in diminishing returns and eventual performance degradation, likely due to the model overfitting to synthetic acoustic artifacts. Notably, the Moderate group remains an outlier, with the zero-shot baseline (23.81\% WER) outperforming all fine-tuned configurations, suggesting that synthetic augmentation may introduce a distribution shift that interferes with the high-quality features that Whisper already utilizes for higher-intelligibility speech. 

\vspace{-5pt}

\subsection{Cross-Corpus Generalization}
Cross-corpus evaluation on the SAP-1102 dataset demonstrates that Clone FT generalizes effectively beyond the source domain, particularly for the CP cohort, where it achieves a 41.6\% WER, a substantial improvement over both the 54.7\% zero-shot baseline and the 48.9\% achieved by Real FT. This success is primarily attributed to the source distribution of the TORGO training set, which is predominantly composed of speakers with CP (7 out of 8). By synthesizing speech that preserves these specific acoustic-phonetic profiles, the model learns generalizable CP-specific deviations that transfer robustly to new subjects. Interestingly, while Real FT shows negligible overall gains over the zero-shot baseline (14.4\% vs. 14.5\%), Clone FT reduces the overall WER to 12.8\%, suggesting that the synthetic data provides a more resilient and lexically diverse training signal that helps bridge the domain gap between datasets. For the ALS cohort, while Real FT marginally outperforms Clone FT  (15.3\% vs. 15.7\%), Clone FT achieves the best performance for the PD cohort at 9.9\% WER. These trends, alongside the significant gains in the CP cohort, suggest that while zero-shot cloning provides a strong regularizing effect across etiologies, its magnitude is directly influenced by the representation of specific pathologies in the source material.

  \vspace{-8pt}
\begin{figure}[h]
    \centering
    \includegraphics[width=\linewidth]{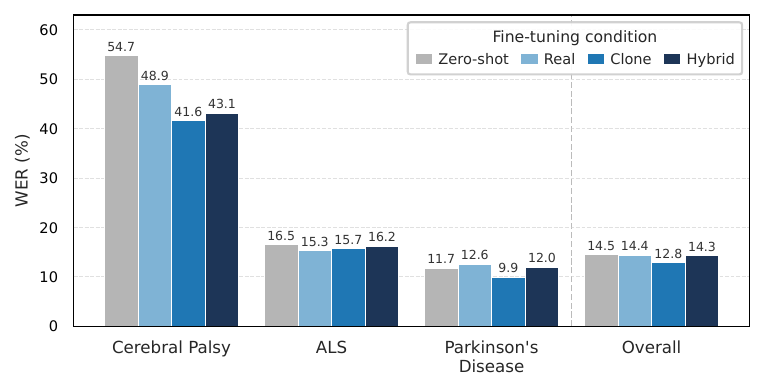}
    \vspace{-20pt}
    \caption{Cross-corpus generalization results (WER \%) on SAP-1102 dataset across three etiologies.}
    \label{fig:sap-wer}
    \vspace{-10pt}
\end{figure}

\vspace{-8pt}
\section{Conclusion}
This study demonstrates that zero-shot voice cloning from a single reference can provide a robust training signal for dysarthric ASR, significantly reducing the data-collection burden on speech-impaired individuals. Our results reveal that performance gains are non-monotonic, with an optimal synthetic volume near 15 hours, beyond which synthesis artifacts may impede generalization. We find that the efficacy of cloned augmentation is highly severity-dependent: while it provides a critical phonetic scaffold for severe dysarthria, it can degrade performance in more intelligible cases. Speaker-embedding analyses further elucidate these speaker-dependent failure modes. Finally, our cross-corpus results on SAP-1102 indicate that cloned dysarthric speech supports broad categorical transfer rather than simple in-distribution adaptation. Future work should explore severity-aware data weighting and adaptive synthesis to further refine these personalized ASR systems.

\section{Generative AI Use Disclosure}
The authors did not use any generative AI in the conceptualization, writing, or preparation of this manuscript.
\bibliographystyle{IEEEtran}
\bibliography{mybib}

\end{document}